\documentclass{article}
\normalsize

\begin{document}
\begin{center}
{\Large\bf A note on Schwarzschild black hole thermodynamics 
\\ in a magnetic universe}
\vspace{0.6cm}
\\
Eugen Radu
\\
{\small
\it Albert-Ludwigs-Universit\"at Freiburg, 
\\Fakult\"at f\"ur Mathematik und Physik,
Physikalisches Institut,
\\Hermann-Herder-Stra\ss e 3, D-79104 Freiburg, Germany
}
\end{center}
\begin{abstract}
We prove that the thermodynamic properties of a Schwarzschild
black hole are unaffected
by an external magnetic field passing through it.
Apart from  the background substraction prescription, this result 
is obtained also by using a counterterm method.
\end{abstract}

The black holes created in astrophysical processes are expected to be well described by 
asymptotically flat solutions of Einstein equations.
However, there is also a  great interest in black holes 
with other kind of asymptotic infinities. 
It particular, it is of interest to determine the effects
that occur when black holes are placed in an external background field, extending to infinity. 

Most of the work on this subject, including pair creation of charged black holes 
has been done in a Einstein-Maxwell theory and in a generalization
of this theory which include a dilaton $\Phi$, whose action is
\begin{eqnarray}
\label{action}
I=\frac{1}{16 \pi} \int_V d^4 x \sqrt{-g} \left(-R+2(\nabla \Phi)^2
+e^{-2a\Phi} F^2 \right)
-\frac{1}{8 \pi} \int_{\partial V} d^3x \sqrt{-h} K,
\end{eqnarray}
where $R$ is the scalar curvature, $F_{\mu \nu}$ is the Maxwell field, and $K$ is 
the trace of the extrinsec curvature of the boundary.
For $a=0$ this is the standard Einstein-Maxwell theory coupled to a massless sclar field. By the no-hair theorems,
$\Phi$ must be constant for solutions describing static black holes.
For $a=1$, (\ref{action}) is a consistent truncation of the low-energy string theory action.
The value $a=\sqrt{3}$ corresponds to standard Kaluza-Klein theory.

An exact solution of Einstein-Maxwell equations describing a black hole
in an background magnetic universe was constructed about 
thirty years ago by Ernst \cite{Ernst1}.
The Melvin magnetic universe is a regular and static, cylindrically symmetric solution
to Einstein-Maxwell theory describing
a bundle of magnetic flux lines in gravitational-magnetostatic equilibrium \cite{Melvin:1963qx}.
This solution has a number of interesting features,
providing the closest approximation in general relativity for an uniform magnetic field.
There exist a fairly extensive literature on the properties of 
black holes in a magnetic universe
(see \cite{KaBu} for a review and relevant references).
However, the general rotating configuration is  very cumbersome to handle
without approximations.

Given the experience with other static nonasymptotically flat black holes,
one may expect that a thermodynamic description of these configurations will include  
also  the value of the background magnetic field as a further parameter.
However, we prove here that this is not the case for a static solution
and a background magnetic field will only distort the geometry of a Schwarzschild solution
without affecting the thermodynamics. 

In spherical coordinates the static electrovacuum  solution found by Ernst reads \cite{Ernst1}
\begin{eqnarray}
\label{metric}
ds^2=\Lambda^2(\frac{dr^2}{f(r)}+r^2 d\theta^2 -f(r)dt^2)
+\frac{r^2 \sin^2 \theta }{\Lambda^{2}}d \varphi^2,
\end{eqnarray}
where 
$\Lambda=1+\frac{B_0^2}{4}r^2 \sin^2 \theta,~f=1-2M/r$,
and the only nonvanishing component of the potential vector $A_i$ in a regular gauge is  
\begin{eqnarray}
\label{A}
A_{\varphi}=\frac{B_0 r^2 \sin^2\theta}{2\Lambda}.
\end{eqnarray}
For $M \to 0$, this solution reduces to Melvin's magnetic geon \cite{Melvin:1963qx}.
The parameter $B_0$ is just the asymptotical cosmological field strength.

We can easily see that this solution  is, 
in terms of the usual definitions, a black hole,
with an event horizon and trapped surfaces.
Although this spacetime is not asymptotically flat, it may represent
an approximation to physical reality in near-zone vicinity of the black hole.
For $r \to \infty$, the line element (\ref{metric}) approaches the Melvin form.
This seems also to constitute the only way to add an external magnetic field to a black hole
without destroying the nonsingular nature of the event horizon \cite{Hiscock:1980zf}.

The event horizon  is at $r=r_0=2M$ and is nonsingular, 
as can be seen by computing
the invariants of the curvature tensor.
It is evident that standard Kruskal coordinates may be introduced 
in order to extend the solution across the event horizon.
The only singularity occurs at $r=0$.

In this model the external field is capable of distorting the geometry from spherical 
symmetry.
The magnetic field has the effect of elongating the event horizon
into a cigaret-shaped object, the long axis being parallel with the magnetic 
field lines \cite{Wild}.
The magnetic field lines remain perpendicular to all points on the event horizon,
analogous to electric lines of force about a conductor.
By applying the Gauss-Bonnet theorem, 
it can also be shown that the surface of the event
horizon is topologically a two-sphere \cite{Wild}.
The surface area of the event horizon is easily evaluated to be $A_H=4\pi r_0^2$,
and remains constant for any value of $B_0$.

The instanton that enters the calculation of the gravitational action is obtained by setting
$\tau=it$ in (\ref{metric}).
Requiring the regularity of the metric at the horizon, 
we find that the Hawking temperature is 
the same as for the Schwarzschild solution, $T_H=1/\beta=1/8\pi M$.
The same result can be obtained by direct calculation of the surface gravity.

Accordingly to Gibbons and Hawking \cite{Gibbons:1976ue}, thermodynamic functions including 
the entropy can be computed directly from 
the saddle point approximation to the gravitational partition function 
(namely the generating functional analytically 
continued to the Euclidean spacetime).

In the semiclassical approximation, the dominant 
 contribution to the path integral will 
come from the neighborhood of saddle points of the action, that is, of classical 
solution; the zeroth order contribution to 
$\log Z$ will be $-I_E$.
The Gibbons-Hawking  surface term evaluated for some $r$ reads
\begin{eqnarray}
I_{b}=
-\frac{1}{8 \pi} \int dt d\theta d\varphi~\sin \theta ~ 
\big( 2rf +\frac{r^2f'}{2}+r^2 f\frac{\Lambda'}{\Lambda}\big),
\end{eqnarray}
where the prime denotes derivative with respect to $r$. 
The volume integral of $R$ is zero by the field equations. 
The volume integral of the Maxwell Lagrangian $F^2$ is not zero, 
but it can be converted to a surface term 
\begin{eqnarray}
\label{Iem}
I_{em}=\frac{1}{8 \pi} \int_{\partial V} d^3x \sqrt{h} A_{\mu}F^{\mu \nu}n_{\nu}= 
\frac{1}{8 \pi} \int dt d\theta d\varphi~\sin \theta ~2rf  (1-\frac{1}{\Lambda}),
\end{eqnarray}
where $n^{\nu}$ is a unit outward pointing to the boundary.
As expected, the sum of these two terms evaluated at infinity diverges.
In the traditional Euclidean path integral approach 
to black hole thermodynamics \cite{Gibbons:1976ue}, 
one has to choose a suitable reference background
and substract it in order to get a finite Euclidean action of black holes.

For our case, the natural background is the Melvin solution. 
However, in this case, apart from  the substraction of $K_{0}$ factor in 
Gibbons-Hawking term we need to substract a further quantity corresponding to 
electromagnetic Melvin contribution.
The sum of these two quantities evaluated for some $r$ is
\begin{eqnarray}
\label{I0}
I_{0}=
\frac{1}{8 \pi} \int dt d\theta d\varphi~\sin \theta 
\left( 2r \sqrt{f} (1-\frac{1}{\Lambda})(\sqrt{f}-1)\right).
\end{eqnarray}
After performing the substraction, the finite result is
\begin{eqnarray}
\label{I}
I_E =\frac{\beta M}{2}.
\end{eqnarray} 
It follows directly that the thermodynamic properties of a Schwarzschild black hole
are not affected by the background magnetic field.
In particular the entropy is $S= A_H/4$ as required.

However, this is a characteristic of the static configurations only.
A tedious but standard calculation confirms that the thermodynamics of rotating solutions
depends nontrivially on the parameter $B_0$.
Heuristically, this is due to the fact that, in the static case, the mass-point source of the 
black hole does not interact directly with the background magnetic field.
This interaction occurs when placing a nonzero electric charge on a black hole.
As discussed in \cite{Hiscock:1980zf} this leads to frame dragging effects and there is no way
to adjust the solution parameters so as to yield a static configuration.
A discussion of the rotating black hole thermodynamics will be presented elsewhere.

The above used reference action substraction procedure is 
generally ill-defined and often leads to confusions
and ambiguities.
A different approach 
has been proposed recently for asymptotically anti-de Sitter spacetimes.
As discussed in \cite{Balasubramanian:1999re}, 
by adding suitable coordinate invariant, boundary 
surface counterterms to the gravitational action, one can obtain 
a well-defined boundary stress-energy tensor and a finite 
Euclidean action for the black hole spacetimes.

So far, however, most of the work on this subject have been restricted
to asymptotically anti-de Sitter spacetimes and their asymptotically flat limit.
It is of interest to apply the counterterm method to spacetimes that are asymptotically Melvin.
For the Schwarzschild-Melvin solution, the simplest counterterm choice is  
\begin{eqnarray}
\label{Sct}
I_{ct}=
\frac{1}{8 \pi} \int d^3 x \sqrt{h} \sqrt{2\mathcal{R}},
\end{eqnarray}
where $\mathcal{R}$ is the Ricci scalar of the boundary metric.
Note  that the same counterterm has been used in \cite{Mann:1999pc,Lau:1999dp}
to regularize the action of asymptotically flat spacetimes with a boundary 
$S_1 \times S_2$.

We find that this prescription removes the divergences and gives a finite action 
that agrees with the reference
spacetime procedure.

Now let us consider the generalization of this result to theories with a dilaton field $\Phi$.
It is a straightforward matter to generate the dilaton version of the solution (\ref{metric}),
by using a dilatonic Harrison transformation found by Dowker and co-workers \cite{Dowker:bt}.
The resulting metric is
\begin{eqnarray}
\label{MSD}
ds^2=\Lambda^{\frac{2}{1+a^2}}(\frac{dr^2}{f(r)}+r^2 d\theta^2 -f(r)dt^2)
+\Lambda^{-\frac{2}{1+a^2}}r^2 \sin^2 \theta d \varphi^2
\end{eqnarray}
where 
\begin{eqnarray}
\Lambda=1+(\frac{1+a^2}{4})B_0^2 r^2 \sin^2 \theta,~~e^{-2a\Phi}=\Lambda^{\frac{2a^2}{1+a^2}}
,~~A_{\varphi}=\frac{2}{(1+a^2)B_0}(1-\frac{1}{\Lambda}). 
\end{eqnarray}

As proven in \cite{Dowker:up}, the action of any solution of this theory can be recasted
as a boundary term
\begin{eqnarray}
\label{action2}
I=
-\frac{1}{8 \pi} \int_{\partial V} d^3x \sqrt{-h} 
e^{-\frac{\Phi}{a}}\nabla_{\mu} (e^{\frac{\Phi}{a}}n^{\mu}).
\end{eqnarray}
From (\ref{MSD}), we notice that the inclusion of a dilaton does not change the
value of Hawking temperature, neither the event horizon area.

A direct computation reveals that the reference action substraction
and the counterterm method give the same value (\ref{I}) for the euclidean 
action and the parameters $a,B_0$ does not appear in the final results.

It would be interesting to find a better understanding of this fact, preferably in terms 
of a microscopic description of black hole thermodynamics.
\\
\\
{\bf Acknowledgement}
\newline
 This work was performed in the context of the
Graduiertenkolleg of the Deutsche Forschungsgemeinschaft (DFG):
Nichtlineare Differentialgleichungen: Modellierung,Theorie, Numerik, Visualisierung.
\newpage

\end{document}